\documentstyle[twocolumn,aps,prl]{revtex}   %PRETTY GALLEY 1/4

\input epsf

\begin{document}
%%%\draft

\twocolumn[\hsize\textwidth\columnwidth\hsize\csname %PRETTY GALLEY 2/4
@twocolumnfalse\endcsname                            %PRETTY GALLEY 3/4

\preprint{}

\title{Temporal and spatial persistence of combustion fronts}

\author{J. Merikoski, J. Maunuksela, M. Myllys, and J. Timonen}
\address{Department of Physics, University of Jyv\"askyl\"a,  
        P.O. Box 35, FIN--40351 Jyv\"askyl\"a, Finland}
\author{M. J. Alava}
\address{Laboratory of Physics, Helsinki University of Technology, 
        P.O. Box 1100, FIN--02015 HUT, Espoo, Finland}

\date{3 December 2002, to appear in PRL}

\maketitle

\begin{abstract}
The spatial and temporal persistence, or first-return distributions
are measured for slow-combustion fronts in paper. 
The stationary temporal and (perhaps less convincingly) spatial 
persistence exponents agree with the predictions based on the front 
dynamics, which asymptotically belongs to the KPZ universality class. 
The stationary short-range and the transient behavior of the 
fronts are non-Markovian, and the observed persistence properties 
thus do not agree with the predictions based on Markovian theory. 
This deviation is a consequence of additional time and length scales, 
related to the crossovers to the asymptotic coarse-grained behavior.
\end{abstract}

\vskip2pc]                                       %PRETTY GALLEY 4 of 4

%%%%%%%%%%%%%%%%%%%%%%%%%%%%%%%%%%%%%%%%%%%%%%%%%%%%%%%%%%%%%%%%%%%%%%%%%%%
%%%%%%%%%%%%%%%%%%%%%%%%%%%%%%%%%%%%%%%%%%%%%%%%%%%%%%%%%%%%%%%%%%%%%%%%%%%

Nonequilibrium systems present a wide range of questions. 
One can try to define necessary conditions for the existence of 
a stationary state and then proceed to a characterization of that state. 
If the dynamics is of an inherently transient character, 
more complicated scenarios arise.
Unlike in the stationary state, in which the behavior of the 
order parameter is described by some probability distribution, 
one encounters concepts such as aging 
and coarsening which reflect the correlations \cite{Bray94}.

A particularly concise and topical approach in nonequilibrium dynamics 
is {\em persistence}, often defined as the probability $P(t)$ that, at a 
point in space, a fluctuating nonequilibrium field (such as a diffusion
field) does not change sign upto time $t$ \cite{Majumdar99,Majumdar96a}. 
This probability may decay algebraically, $P(t) \sim t^{-\theta}$, 
with a persistence exponent $\theta$. 
In some cases $\theta$ may be unrelated 
to any of the other exponents that characterize the system, 
in particular those describing the critical decay of the 
temporal and spatial correlation functions. 
Of special interest in this respect are {\em non-Markovian} random processes,
for the extra challenges they present.

Due to the statistical nature of the problem, 
clear experimental demonstrations of persistence are still 
quite rare \cite{Wong01,Yurke97,Tam97,Dougherty02}.
In the present Letter, we consider persistence as observed in the 
experiments on the propagation of slow-combustion fronts in paper sheets. 
We have earlier \cite{Maunuksela97,Myllys00,Myllys01} shown that 
at long spatial and temporal scales the dynamics of these fronts follows
the universality class of the 1+1-dimensional Kardar-Parisi-Zhang (KPZ) 
equation \cite{Kardar86,Halpin-Healy95}
\begin{equation}
\label{kpz}
\frac{\partial h}{\partial t} = \nu \nabla^2 h + 
\frac{\lambda}{2} (\nabla h)^2 + \eta , 
\end{equation}
where $h=h(x,t)$ is the height of the interface at point $x$ 
and time $t$, $\nu$ is the surface tension parameter, 
$\lambda$ is the strength of the nonlinearity, 
and $\eta=\eta(x,t)$ is Gaussian white noise. 
We have also shown\cite{Myllys00,Myllys01} that at short scales the 
short-range correlations in the effective noise prevent the system from 
displaying true scaling behavior.

We consider here the first-return properties of the experimentally observed 
fronts, $h(x,t)$, in spatial and temporal domains, at both stationary and 
transient stages, and compare them with various theoretical predictions. 
For short time and length scales, the correlations in the noise in 
our experiments are relevant, and the dynamics is non-Markovian. 
We study the implications of this on persistence.

\begin{figure}[htb]
% \vspace{80truemm}
\centerline{\epsfxsize=0.6\columnwidth\epsfbox{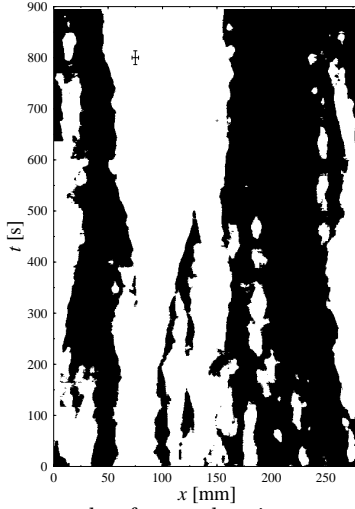}}
% \null\vfill
\caption{An example of a combustion experiment. 
In the height-fluctuation field the black color 
indicates fluctuations in the positive and white in the negative direction. 
The cross-over scales $\tau_c$ and $\ell_c$, 
see Figs.~\ref{FigStatTemp}(b) and \ref{FigStatSpat}(b), are marked by bars. 
\label{FigDemo}}
\end{figure}

A complete description of our experimental setup is given 
in Ref.~\cite{Myllys01}. 
We have chosen to analyze data for the 80 gm$^{-2}$ copier paper 
as a typical case, and also since it presents the best statistics. 
In Fig.~\ref{FigDemo} we show the height-fluctuation field observed 
in a typical sample ``burn''.

%The theoretical expectations are as follows: 

In the absence of nonlinearity ($\lambda=0$), the long-range dynamics 
would be described by a linear diffusion equation with thermal 
noise, the Edwards-Wilkinson (EW) equation. 
This case was analyzed in the temporal domain by Krug {\em et al.} in the 
context of linear growth equations \cite{Krug97} and by Bray and 
Majumdar in terms of the spatial return characteristics \cite{Majumdar01a}. 
In the {\em stationary state} one finds $\theta_s=1/2$, which for 
these linear equations is the same as for the transient behavior. 

For $\lambda\neq 0$ the stochastic process, the KPZ height 
fluctuations, is non-Gaussian, and the stationary and transient 
temporal persistences are theoretically difficult to analyze. 
Kallabis and Krug observed, starting from a numerical growth model, 
that the persistence behavior can be characterized by considering 
the scaling functions of the general persistence probability $P(t_0,t)$, 
where $t$ is measured beginning from time $t_0$ after the start of the 
kinetics from a flat initial profile \cite{Kallabis99}. 
This has two limiting behaviors, the {\it transient behavior} 
for $t_0$ before saturation and the {\it stationary-state behavior} 
for $t_0$ after it, thus defining two persistence exponents. 
The numerical results \cite{Kallabis99} follow the general {\em conjecture}
that for the stationary state 
\begin{equation} 
\theta_s = 1 - \beta, 
\label{conject}
\end{equation}
where $\beta$ describes the temporal two-point correlation function of 
the interface fluctuations (to be defined below), which 
can be justified by the self-affine nature of the process. 
For the transient regime, Kallabis and Krug observed, e.g., that persistence 
{\em depends on the up-down asymmetry of the dynamics}, in their model with 
values 1.2 and 1.6 for the transient temporal persistence exponents 
for fluctuations in the up and down directions, 
respectively \cite{Kallabis99}. 
In the stationary state the asymmetry was reflected only  
as different normalizations of the persistence probability 
for fluctuations in the up and down directions.

One may likewise consider any particular front at a fixed
time $t$ and look at the interface profile as a stochastic
process as discussed by Bray and Majumdar
\cite{Majumdar01a}, who studied first-passage properties in space. 
In analogy with the temporal case, the probability that the interface 
stays above or below a given reference level may have a power-law decay.
Moreover, depending on whether the configuration represents
the stationary state or whether it has essentially finite
fluctuations, one again has two different exponents. 
In terms of the interface morphology, they can be related to 
the roughness exponent $\chi$ (see below).

To characterize persistence, we consider $f^{\rm temp}_{\pm}(\tau)$, 
the first-return distributions, {\it i.e.}, the distributions of return 
time $\tau$, defined as the time a variable stays above ($+$) or 
below ($-$) a given reference level. 
The persistence exponents $\theta^{\rm temp}_\pm$ 
and $\theta^{\rm spat}_\pm$ describe the decay of 
the related temporal and spatial persistence probabilities, 
respectively, and are defined via \cite{Krug97,Majumdar01a} 
\begin{eqnarray}
 \label{EqExp}
  P^{\rm temp}_\pm(\tau) \sim \tau^{-\theta^{\rm temp}_\pm}
 \ \ \ \ {\rm and} \ \ \ \
  P^{\rm spat}_\pm(\ell) \sim \ell^{-\theta^{\rm spat}_\pm},
\end{eqnarray}
where $\tau$ and $\ell$ denote the persistent time- and lengthscales.
In the transient regime these quantities are computed by
sampling within a suitably chosen time window (see below),
and denoted by a tilde, e.g., $\tilde{\theta}^{\rm temp}_\pm$.

Since a combustion front propagates with a finite average 
velocity $v$, we look at the fluctuations $\delta h(x,t)$,
at a fixed point $x=x_0$ and time $t$,
in the front $h(x,t)$ around its average height, 
\begin{equation}
  \label{Eqdh}
  \delta h(x,t) \equiv h(x,t) -\overline{h}(t).
\end{equation}
Here overbar denotes a spatial average at time $t$. 
For practical purposes we define the return times as follows: 
$\tau_{+}$ is the length of the time interval 
between $t_1$ and $t_2$ such that 
$\delta h(x_0,t_1)=0=\delta h(x_0,t_2)$, 
and $\delta h(x_0,t)>0$ for all $t\in\,]t_1,t_2[$. 
The return time $\tau_{-}$ is defined analogously 
for $\delta h(x_0,t)<0$.
For discrete sampling times we determine the crossing times 
by using linear interpolation.

The temporal persistence probabilities \cite{Krug97} are related to 
the corresponding first-return distribution via 
\begin{eqnarray}
  \label{EqPtau}
  P^{\rm temp}_{\pm}(\tau)\equiv P(\tau_{\pm}\!\ge\tau) 
            = 1-\int_{-\infty}^{\tau} f^{\rm temp}_{\pm}(\tau')\,d\tau'\ .
\end{eqnarray}
As explained in Ref.~\cite{Majumdar01b}, in discrete time 
(and also in discrete space) sampling, one misses very short excursions, 
and correct normalization of $P^{\rm temp}_\pm$ and $P^{\rm spat}_\pm$ 
is difficult. 
Therefore, we prefer to use the distributions $f^{\rm temp}_\pm$ and 
$f^{\rm spat}_\pm$, instead of their integrals, 
for the determination of the persistence exponents. 
In the limit of long time and length scales ($\tau$ and $\ell$), 
the problems in these functions due to discrete sampling 
should disappear.

The corresponding spatial quantities 
\cite{Majumdar01a} at fixed times are defined analogously. 
$P^{\rm spat}_\pm(\ell)$ is the probability that the front 
stays above ($+$) or below ($-$) the reference level over a 
distance $\ell$, and $f^{\rm spat}_\pm(\ell)$ denotes the 
corresponding return-length distribution.

The conjecture of Eq.~(\ref{conject}) for the temporal persistence,
and the arguments of Bray and Majumdar, make it necessary to outline
the behavior of the two-point correlation functions. 
These, and the associated critical exponents are defined
in the usual way as \cite{Halpin-Healy95}
\begin{eqnarray}
 \label{EqCor}
  C^{\rm temp}(\tau) 
   = \langle \overline{|\delta h(x,t)-\delta h(x,t+\tau)|}\rangle 
   \sim \tau^{\beta}
  \\
  C^{\rm spat}(\ell) 
   = \langle \overline{|\delta h(x,t)-\delta h(x+\ell,t)|}\rangle 
   \sim \ell^{\chi},
\end{eqnarray}
in which the overbar denotes spatial and the brackets disorder 
averaging \cite{eivahennetty}. In our earlier works, the long-range 
scaling of the correlation functions was shown to be described by 
the KPZ asymptotics with $\chi \simeq 1/2$ and $\beta \simeq 1/3$ 
\cite{Maunuksela97,Myllys00,Myllys01}.

\begin{figure}[htb]
 \null
\centerline{\epsfxsize=0.55\columnwidth\epsfbox{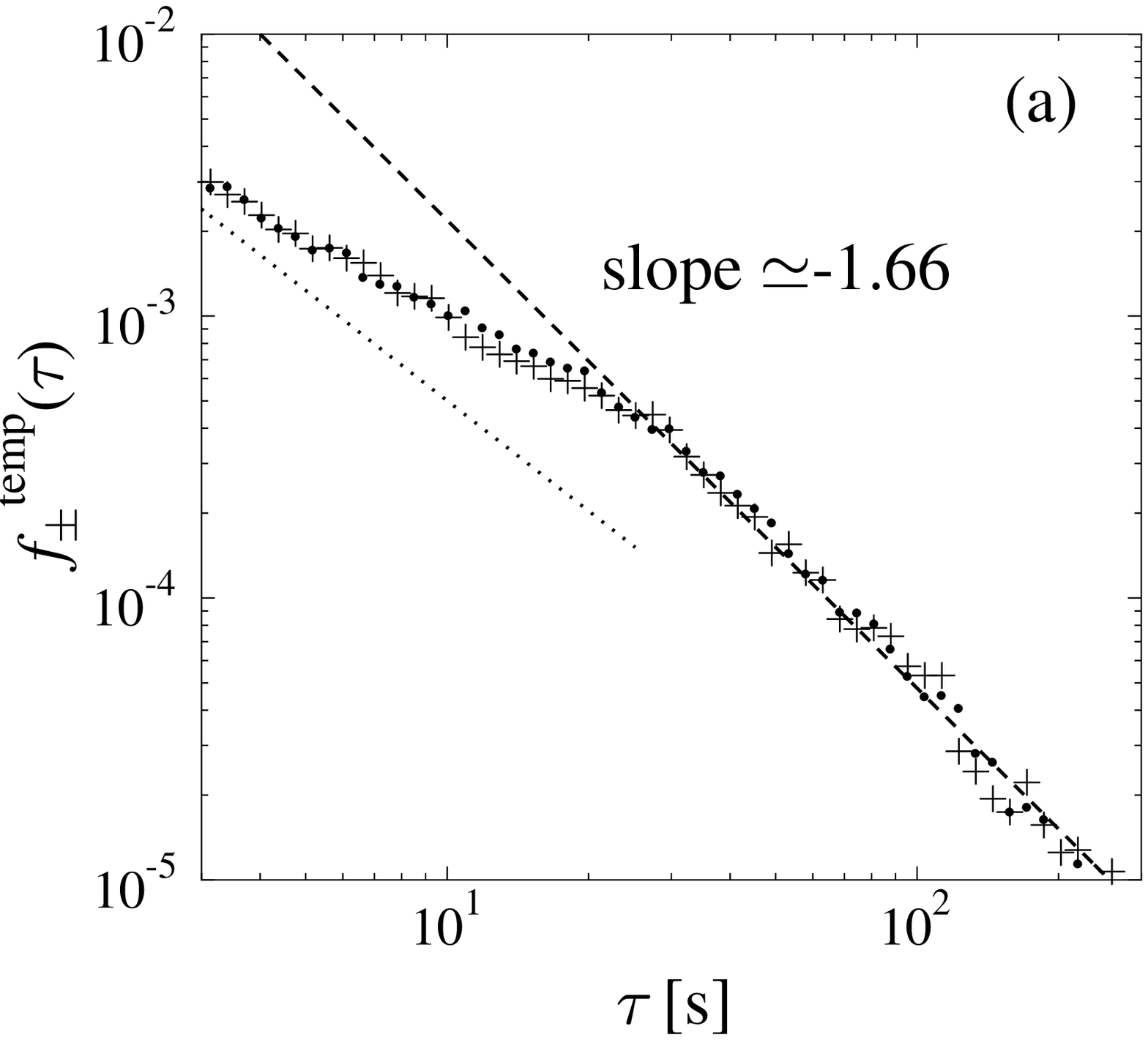}}
\centerline{\epsfxsize=0.55\columnwidth\epsfbox{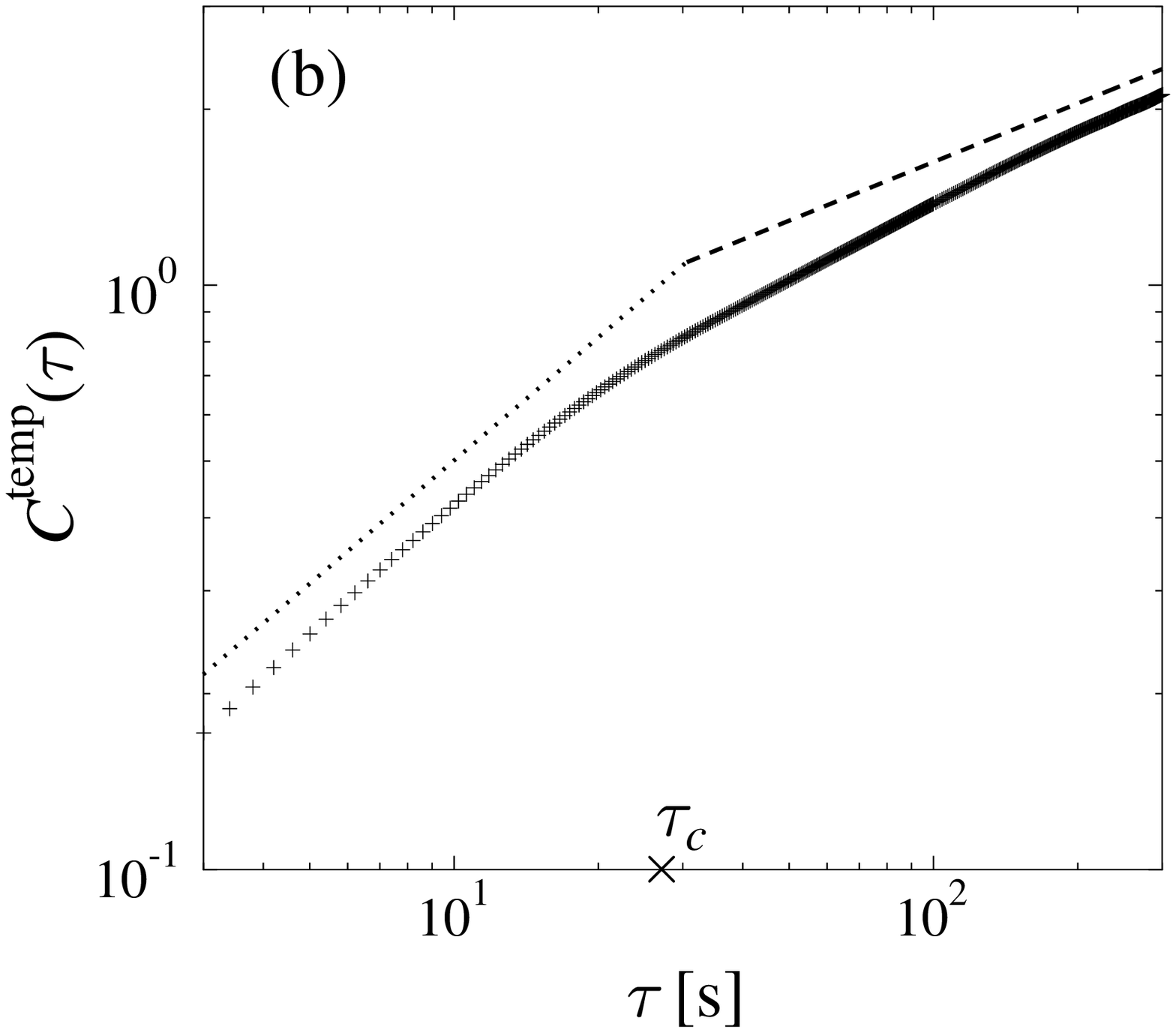}}
% \vspace{80truemm}
 \vfill
 \caption{The stationary temporal 
  (a) first-return distributions $f^{\rm temp}_{+}(\tau)$ 
  and $f^{\rm temp}_{-}(\tau)$ denoted by crosses and dots, respectively, 
  and (b) correlation function $C^{\rm temp}(\tau)$. 
  In all figures, the distributions are normalized to unity and 
  the horizontal axis is logarithmically binned. 
  The dashed and the dotted lines correspond to the asymptotic KPZ 
  and apparent short-range values \protect\cite{Myllys01} of $\beta$, 
  in Fig.~(a) via the conjecture $f(\tau)\sim \tau^{-(2-\beta)}$ 
  from Eq.~(\protect\ref{conject}). 
 \protect\label{FigStatTemp}}
\end{figure}

In Fig.~\ref{FigStatTemp} we show our experimental results for temporal 
persistence in the stationary state. Above a cross-over scale $\tau_c$
of the order of 10 seconds (corresponding to fronts that have
propagated, on the average, by about 45 mm, see Fig.~\ref{FigDemo}
again), this figure
indicates agreement with the theoretical expectation,
$ \theta^{\rm temp}_\pm = 1-\beta $. 
The dashed line in Fig.~\ref{FigStatTemp}(a) follows Eq.~(\ref{conject}). 
Below $\tau_c$ the data in Fig.~\ref{FigStatTemp}(a) show no real scaling 
regime, in analogy to the curvature visible in $C_{\pm}^{\rm temp}(\tau)$ 
shown Fig.~\ref{FigStatTemp}(b). 
We find no difference between the first-return 
distributions in the positive and negative 
directions since our distributions are separately normalized to 
one, so the anisotropy does not appear in the plots.

\begin{figure}[htb]
 \null
\centerline{\epsfxsize=0.55\columnwidth\epsfbox{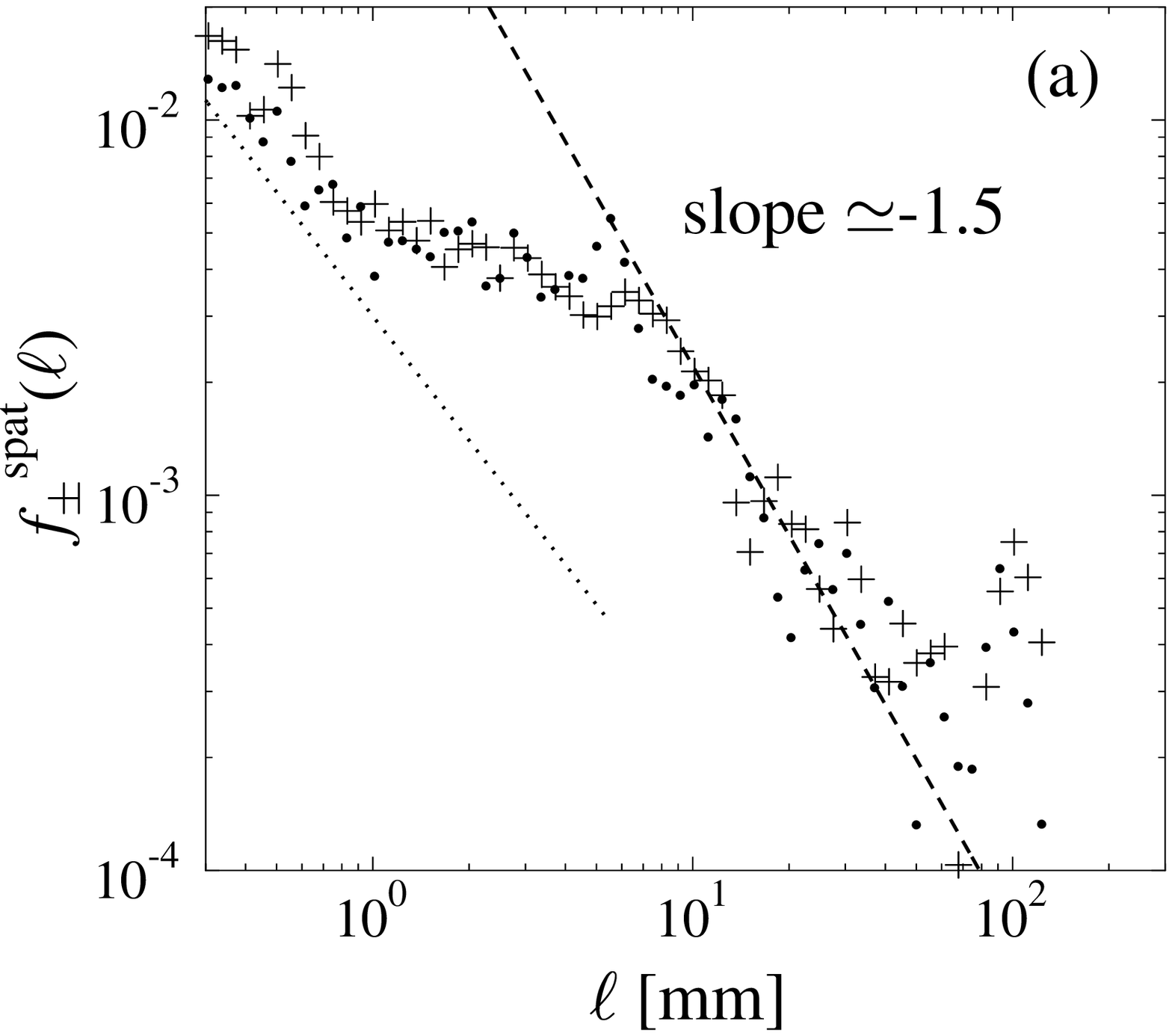}}
\centerline{\epsfxsize=0.55\columnwidth\epsfbox{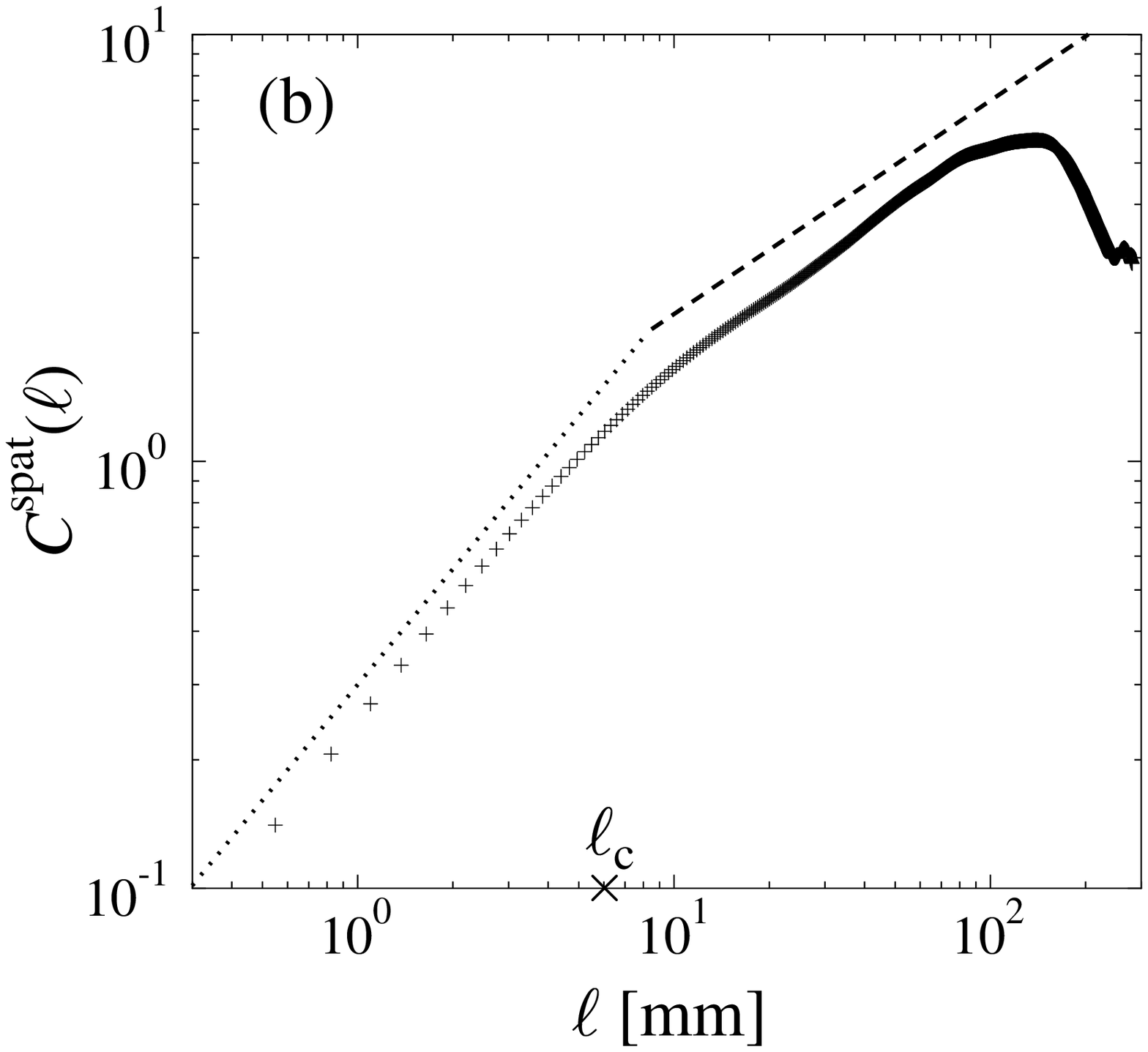}}
% \vspace{80truemm}
 \vfill
 \caption{The stationary spatial 
  (a) first-return distributions $f^{\rm spat}_\pm(\ell)$ 
  denoted by crosses and dots, respectively, 
  and (b) correlation function $C^{\rm spat}(\ell)$. 
  The dashed and the dotted lines correspond to the asymptotic KPZ 
  and apparent short-range values \protect\cite{Myllys01} of $\chi$, 
  in Fig.~(a) via the conjecture $f(\ell)\sim \ell^{-(2-\chi)}$ from 
  Eq.~(\protect\ref{conject}). 
 \protect\label{FigStatSpat}}
\end{figure}

The measured first-return distributions for spatial persistence
in the stationary state are shown in Fig.~\ref{FigStatSpat}.
Now the asymptotic behavior is roughly consistent with 
$ \theta^{\rm spat}_\pm = 1-\chi $. 
The rather long crossover regions \cite{Myllys01} in the correlation 
functions $C(\tau)$ and $C(\ell)$ in Figs.~\ref{FigStatTemp}(b) 
and \ref{FigStatSpat}(b), which precede the long-range regimes, are
directly visible in the persistence data as well. The crossover is 
here located at about 10 mm, which is less than the scale 
$\tau_c / v$ would indicate.

The expected persistence behavior takes place only on long enough 
scales, where the physics is {\em coarse-grained} so as to 
obey the KPZ equation. 
There are correlations in the effective noise (local height 
increments), with decay scales of a few seconds and a few 
millimeters \cite{Myllys00}. 
In the short-range regime of the stationary-state data, 
the spatial and temporal statistics are quite 
far from the scaling conjecture ``$\theta = 1-\beta$''. 
This agrees with the fact that the dynamics becomes Markovian only 
asymptotically. The short-range persistence does not result from
an effectively stationary process that would differ from the long-range
dynamics only by the fact that the two-point exponents are not
defined. The deviation is greatest in the case of the temporal behavior, for 
which persistence decays slower than expected from the 
correlation function.

\begin{figure}[htb]
 \null
\centerline{\epsfxsize=0.55\columnwidth\epsfbox{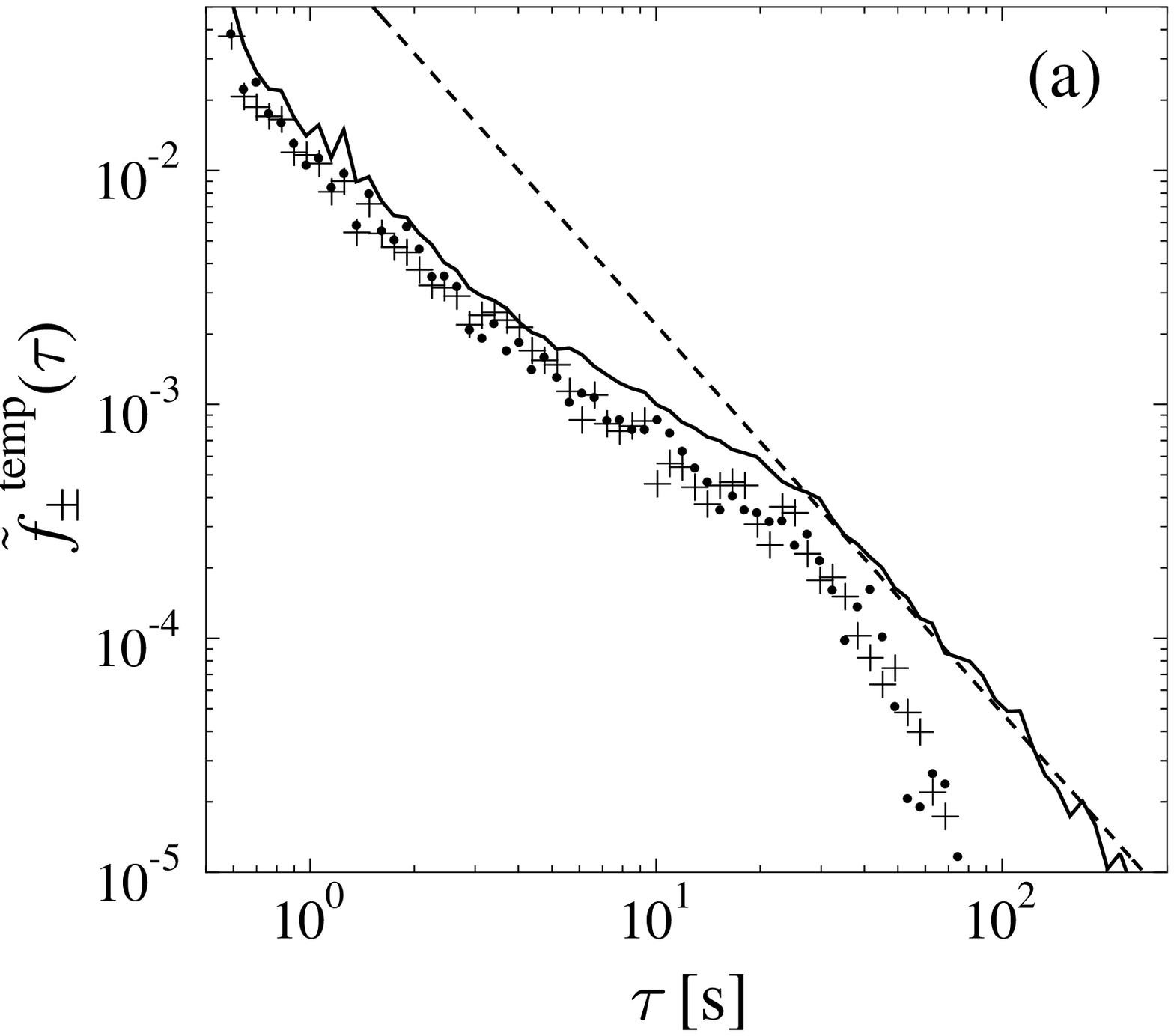}}
\centerline{\epsfxsize=0.55\columnwidth\epsfbox{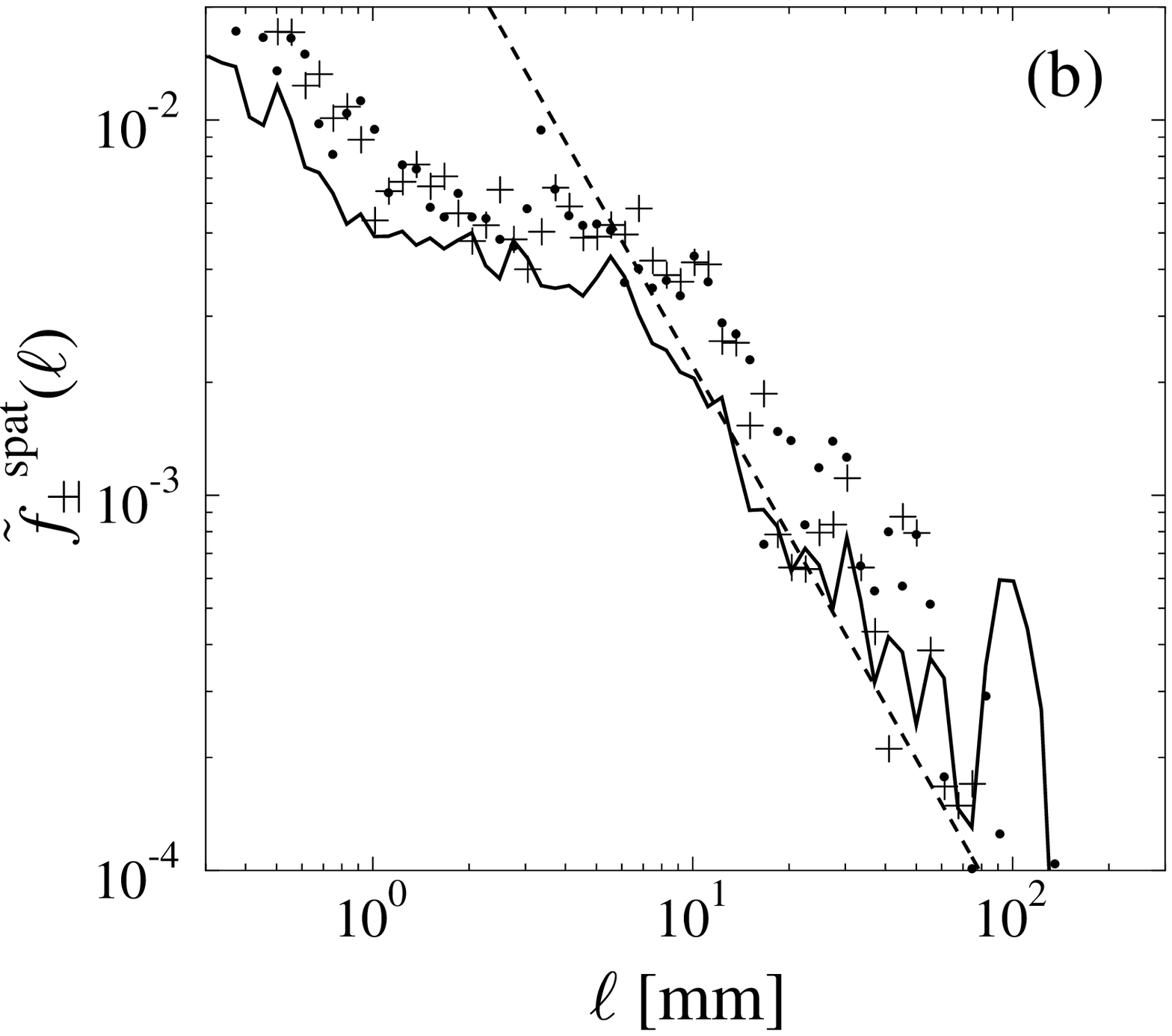}}
% \vspace{80truemm}
 \vfill
 \caption{Transient regime 
  (a) temporal first-return distributions $\tilde{f}^{\rm temp}_\pm(\tau)$ 
  and (b) spatial first-return distributions $\tilde{f}^{\rm spat}_\pm(\ell)$. 
  In both cases crosses denote excursions to positive and dots to 
  negative direction. 
  By the full curves we show the averages (over $\pm$ 
  directions) of the stationary distributions $f^{\rm temp}_\pm$ 
  and $f^{\rm spat}_\pm$ of Figs.~\protect\ref{FigStatTemp} 
  and~\protect\ref{FigStatSpat}, respectively.
 \protect\label{FigTran}}
\end{figure}

The first-return distributions for {\it the transient-time regime} of 
the dynamics are shown in Fig.~\ref{FigTran} for a time window
from the beginning of the burn to just below the saturation time. 
In the temporal case the simulation result for the model 
of Ref. \cite{Kallabis99} 
was $\tilde{\theta}^{\rm temp}_{\pm} \sim 1.2...1.6$ . 
In our temporal transient distribution shown in
Fig~\ref{FigTran}(a), the 
asymptotics beyond the crossover from short-range behavior shows
a steeper decay than in the saturated distribution, but the length
of this asymptotic part is so short that definitive conclusions
cannot be made. It may as well be related to
some kind of cutoff behavior. No indications of an
up-down anisotropy can be seen. 
In the spatial transient distribution the asymptotics is better
defined, and it is interesting to notice that it is given by
$\ell^{-(2-\chi)}$ with $\chi=1/2$ as in the saturated
regime (see Fig~\ref{FigTran}(b)).

The main features of the observed transient 
persistences can be summarized in two findings: First, in our data 
the spatial long-range scaling is reminiscent of the stationary state. 
Second, there is no simple short-range behavior below the 
cross-over scales. In some cases the short-range scalings resemble 
power-law ones, albeit over rather short ranges, but the effective 
persistence exponents are never in agreement with those related to 
the decay of correlations.

In principle, for a time window above the cross-over time $\tau_c$ 
but below the saturation time some ``expected'' (KPZ) temporal
transient behavior could be observed. 
However, the typically wide \cite{Myllys01} cross-over region 
around $\tau_c$ would interfere with it, unless the saturation time 
could be made long enough by, e.g., considerably increasing the system
size. 
Recall that the origin of the temporal transient exponents in KPZ-type 
dynamics, as seen in Ref.~\cite{Kallabis99}, is the asymmetric interface 
(valleys and hilltops). In our case, these features can be 
visible {\em only} for $\tau>\tau_c$ and $\ell>\ell_c$ \cite{Myllys01}.

To summarize, we have studied the persistence 
properties of fluctuating combustion fronts in paper.
Asymptotically both temporal and spatial persistence 
follow the theoretical expectations for the stationary case. 
For the short-range behaviors the KPZ physics is irrelevant, 
and the first-return distributions deviate from those based on a 
stationary Markovian stochastic process \cite{Costa02}. 
Instead, the physics indicates the presence of memory effects.
Our results have theoretical implications also for problems, 
where the KPZ scaling is seen only asymptotically \cite{Sneppen92}.

%%%%%%%%%%%%%%%%%%%%%%%%%%%%%%%%%%%%%%%%%%%%%%%%%%%%%%%%%%%%%%%%%%%%%%%%%%%
%%%%%%%%%%%%%%%%%%%%%%%%%%%%%%%%%%%%%%%%%%%%%%%%%%%%%%%%%%%%%%%%%%%%%%%%%%%

\vskip2truemm

We wish to acknowledge discussions on stochastic dynamics with 
Otto Pulkkinen and on kinetic roughening with Tapio Ala-Nissila.
This work has been in part supported by the Academy of Finland 
under the Center of Excellence Program (Project No. 44875).

%%%%%%%%%%%%%%%%%%%%%%%%%%%%%%%%%%%%%%%%%%%%%%%%%%%%%%%%%%%%%%%%%%%%%%%%%%%
%%%%%%%%%%%%%%%%%%%%%%%%%%%%%%%%%%%%%%%%%%%%%%%%%%%%%%%%%%%%%%%%%%%%%%%%%%%

\eject

%%%%%%%%%%%%%%%%%%%%%%%%%%%%%%%%%%%%%%%%%%%%%%%%%%%%%%%%%%%%%%%%%%%%%%% FIG1

%%%%%%%%%%%%%%%%%%%%%%%%%%%%%%%%%%%%%%%%%%%%%%%%%%%%%%%%%%%%%%%%%%%%%%% FIG2

%\vskip-6truemm

%%%%%%%%%%%%%%%%%%%%%%%%%%%%%%%%%%%%%%%%%%%%%%%%%%%%%%%%%%%%%%%%%%%%%%% FIG3

% \vskip-6truemm

%%%%%%%%%%%%%%%%%%%%%%%%%%%%%%%%%%%%%%%%%%%%%%%%%%%%%%%%%%%%%%%%%%%%%%% FIG4

%%%%%%%%%%%%%%%%%%%%%%%%%%%%%%%%%%%%%%%%%%%%%%%%%%%%%%%%%%%%%%%%%%%% FIGEND


\begin{thebibliography}{100}
\bibitem{eivahennetty} For the sake of comparing
with the first-return distributions one has to use the ``bare''
correlation functions. That is, the intrinsic widths
can not be subtracted \protect{\cite{Myllys01}}.
\end{thebibliography}
\end{document}